\documentclass[12pt,preprint]{aastex}
\usepackage{longtable}

\def\gtorder{\mathrel{\raise.3ex\hbox{$>$}\mkern-14mu
             \lower0.6ex\hbox{$\sim$}}} 
\def\ltsima{$\; \buildrel < \over \sim \;$}
\def\simlt{\lower.5ex\hbox{\ltsima}}
\def\gtsima{$\; \buildrel > \over \sim \;$}
\def\simgt{\lower.5ex\hbox{\gtsima}} 

\begin{document} 


\title{Core-Collapse Supernovae from the Palomar Transient Factory:\\ 
Indications for a Different Population in Dwarf Galaxies}


\author{Iair Arcavi\altaffilmark{1,2}, Avishay Gal-Yam\altaffilmark{1},
Mansi M. Kasliwal\altaffilmark{3}, Robert M. Quimby\altaffilmark{3}, 
Eran O. Ofek\altaffilmark{3}, Shrinivas R. Kulkarni\altaffilmark{3},
Peter E. Nugent\altaffilmark{4}, S. Bradley Cenko\altaffilmark{5}, 
Joshua S. Bloom\altaffilmark{5}, Mark Sullivan\altaffilmark{6},
D. Andrew Howell\altaffilmark{7,8}, Dovi Poznanski\altaffilmark{4,5,9},
Alexei V. Filippenko\altaffilmark{5}, Nicholas Law\altaffilmark{3,10},
Isobel Hook\altaffilmark{6,11}, Jakob J\"onsson\altaffilmark{6}, 
Sarah Blake\altaffilmark{6}, Jeff Cooke\altaffilmark{3}, Richard 
Dekany\altaffilmark{12}, Gustavo Rahmer\altaffilmark{12}, David 
Hale\altaffilmark{12}, Roger Smith\altaffilmark{12}, Jeff 
Zolkower\altaffilmark{12}, Viswa Velur\altaffilmark{12}, Richard 
Walters\altaffilmark{12}, John Henning\altaffilmark{12}, Kahnh 
Bui\altaffilmark{12}, Dan McKenna\altaffilmark{12}, and
Janet Jacobsen\altaffilmark{4}}

\altaffiltext{1}{Benoziyo Center for Astrophysics, Faculty of Physics,
  The Weizmann Institute of Science, Rehovot 76100, Israel.}
\altaffiltext{2}{email iair.arcavi@weizmann.ac.il.}
\altaffiltext{3}{Cahill Center for Astrophysics, California Institute
  of Technology, Pasadena, CA, 91125.}
\altaffiltext{4}{Computational Cosmology Center, Lawrence Berkeley
  National Laboratory, 1 Cyclotron Road, Berkeley, CA 94720.}
\altaffiltext{5}{Department of Astronomy, University of California,
  Berkeley, CA 94720-3411.}
\altaffiltext{6}{Department of Physics (Astrophysics), University of
  Oxford, Keble Road, Oxford, OX1 3RH, UK.}
\altaffiltext{7}{Las Cumbres Observatory Global Telescope Network,
  6740 Cortona Dr., Suite 102, Goleta, CA 93117.}
\altaffiltext{8}{Department of Physics, University of California,
  Santa Barbara, Broida Hall, Mail Code 9530, Santa Barbara, CA 
  93106-9530.}
\altaffiltext{9}{Einstein Fellow.}
\altaffiltext{10}{Dunlap Institute for Astronomy and Astrophysics,
  University of Toronto, 50 St. George Street, Toronto M5S 3H4,
  Ontario, Canada.}
\altaffiltext{11}{INAF -- Osservatorio Astronomico di Roma, via
  Frascati 33, 00040 Monteporzio (RM), Italy.}
\altaffiltext{12}{Caltech Optical Observatories, California Institute
  of Technology, Pasadena, CA 91125.}



\newpage

\begin{abstract} 

We use the first compilation of 72 core-collapse supernovae (SNe) from
the Palomar Transient Factory (PTF) to study their observed subtype
distribution in dwarf galaxies compared to giant galaxies. Our sample
is the largest single-survey, untargeted, spectroscopically
classified, homogeneous collection of core-collapse events ever
assembled, spanning a wide host-galaxy luminosity range (down to $M_r
\approx -14$ mag) and including a substantial fraction ($>20\%$) of
dwarf ($M_r \geq -18$ mag) hosts. We find more core-collapse SNe in
dwarf galaxies than expected and several interesting trends emerge. We
use detailed subclassifications of stripped-envelope core-collapse SNe
and find that all Type I core-collapse events occurring in dwarf
galaxies are either SNe~Ib or broad-lined SNe~Ic (SNe~Ic-BL), while
``normal'' SNe~Ic dominate in giant galaxies. We also see a
significant excess of SNe~IIb in dwarf hosts. We hypothesize that in
lower metallicity hosts, metallicity-driven mass loss is reduced,
allowing massive stars that would have appeared as ``normal'' SNe~Ic
in metal-rich galaxies to retain some He and H, exploding as Ib/IIb
events. At the same time, another mechanism allows some stars to
undergo extensive stripping and explode as SNe~Ic-BL (and presumably
also as long-duration gamma-ray bursts). Our results are still limited
by small-number statistics, and our measurements of the observed
$N(\textrm{Ib/c})/N(\textrm{II})$ number ratio in dwarf and giant
hosts ($0.25_{-0.15}^{+0.3}$ and $0.23_{-0.08}^{+0.11}$, respectively;
1$\sigma$ uncertainties) are consistent with previous studies and
theoretical predictions. As additional PTF data accumulate, more
robust statistical analyses will be possible, allowing the evolution
of massive stars to be probed via the dwarf-galaxy SN population.

\end{abstract} 




\section{Introduction} 

Understanding the formation, evolution, and ultimate fate of massive
stars is a key open question in astrophysics. A particularly
interesting problem concerns the formation of highly stripped objects
known as Wolf-Rayet (W-R) stars. It is still unclear whether W-R stars
are formed through wind driven mass loss from single very massive
stars, or if binary interactions play an important role in which case the
minimal initial mass of W-R stars may be much lower. The final stages
in the evolution of a massive star influence the type of explosion it
will undergo at the end of its life; studying the resulting
core-collapse supernovae (SNe) is thus an attractive way to learn
about these massive progenitors. The hope is to reveal how the
physical properties of the star (such as its mass and metallicity)
determine its pre-explosion mass loss, as reflected in the properties
of the ensuing SN. An additional benefit is that luminous SN
explosions are visible to great distances and hence provide access to
a wide range of stellar populations, while normal individual stars can
be effectively studied only within the Local Group.

Core-collapse SN explosions of massive stars are observationally
classified according to their spectroscopic properties (see Filippenko
1997 for a review). H-rich Type IIP (``plateau'') SNe have been
associated with red supergiant progenitors (Smartt 2009, and
references therein). The progenitors of Type IIb SNe appear to have
lost most (but not all) of their hydrogen envelope before they
exploded. Further mass loss leads to Type Ib explosions showing strong
He lines but lacking H, while SNe~Ic, lacking both H and He, probably
arise from the most highly stripped progenitors. A particular subclass
of SNe~Ic, displaying very broad lines (SNe~Ic-BL) indicating high
expansion velocities, is notable for the fact that its members are the
only events to have been observationally linked with gamma-ray bursts
(GRBs) and X-ray flashes (XRFs; see Woosley \& Bloom 2006 for a
review).  While it stands to reason that H-poor SNe arise from H-poor
W-R stars, attempts to directly detect such stars in pre-explosion
images of the locations of SNe~Ib and Ic have so far failed (Gal-Yam
et al. 2005; Maund et al. 2005; Crockett et al. 2008b; Smartt
2009). The exact identity of the progenitor stars of SNe~IIb/Ib/Ic, as
well as those of GRBs and XRFs, remains an unsolved puzzle.

Given the limited efficacy of direct pre-explosion imaging studies,
one can attempt to study massive-star populations using statistical
analyses of the resulting SNe. Two main methods have been employed for
this purpose thus far, as follows.

The number ratio of SNe~Ib/c to SNe~II as a function of host-galaxy
metallicity has been measured using host luminosity as a metallicity
proxy for 280 and 497 core-collapse SNe (Prantzos \& Boissier 2003;
Boissier \& Prantzos 2009, respectively), as well as via direct
metallicity measurements for 159 core-collapse SNe (Prieto et
al. 2008). Both methods show that the ratio of SNe~Ib/c to SNe~II
increases with increasing host metallicity. This suggests that
wind-driven mass loss plays an important role in the formation of
stripped W-R stars exploding as SNe~Ib/c, and that at lower
metallicities massive stars that would have ended as SNe~Ib/c explode
instead as Type II events (``failed Ib/c''). Heger et al. (2003)
predicted this behavior, but their decline rate for the SN~Ib/c ratio
with decreasing metallicity, as well as that of Fryer et al. (2007),
does not agree with the Boissier \& Prantzos (2009) data.

The location of a SN in its host galaxy can also be indicative of the
mass and metallicity of its progenitor. Anderson \& James (2009)
examined 173 core-collapse SN locations and confirmed previous results
by van den Bergh (1997), Tsvetkov et al. (2004), and Hakobyan (2008)
which indicated that SNe~Ib/c are preferentially found in more central
regions of their host galaxies compared to SNe~II (see also Boissier
\& Prantzos 2009; Hakobyan 2009), with SNe~Ic the most centrally
concentrated. This implies a progenitor-metallicity sequence
(II-Ib-Ic) with SNe~Ic having the highest metallicity
progenitors. Like the population-ratio studies, these results also
suggest that SN~Ib/c progenitor analogs (in terms of mass) would
explode as SNe~II in low-metallicity regions. Other location studies,
however, indicate otherwise. Fruchter et al. (2006) found that
long-duration GRBs occur preferentially in faint (metal-poor) galaxies
(see also Stanek et al. 2006) and in brighter regions than the general
core-collapse population (based on 16 core-collapse SN
locations). Kelly et al. (2008) showed that SNe~Ic (26 in their
sample) occur in areas within their hosts similar to those of
long-duration GRBs. Taken together, these results, as well as those of
Leloudas et al. (2010), would suggest that some SN~Ic progenitor
analogs in low-metallicity environments tend to explode as SNe~Ic-BL
(accompanied by long GRBs) rather than as SNe~II.

While the results from the studies described above are intriguing and
hint at the power of the statistical analysis of large SN samples, a
clear picture has yet to emerge. Perhaps this is due to the small
number of events in dwarf, low-metallicity galaxies (leading Prantzos
\& Boissier 2003 and Boissier \& Prantzos 2009, for example, to extend
the ``dwarf'' galaxy bin to luminosities as bright as $-19$ mag). It
should also be noted that many studies to date were based on
heterogeneous compilations of SN data, inheriting their strong and
unknown selection effects, and many included SNe discovered decades
ago, the exact classification of which was unclear (Anderson \& James
2009). Here we present a population-ratio analysis for the first
sample of 72 core-collapse SNe from the Palomar Transient Factory
(PTF; Rau et al. 2009; Law et al. 2009), 15 of which are in dwarf
galaxies having $r$-band magnitudes $M_r \geq -18$. The observational
data are described in $\S~2$ and the results in $\S~3$. We discuss our
findings in $\S~4$ and summarize in $\S~5$.

\section{Observations}

\subsection{The Palomar Transient Factory}

PTF is a wide-field variability survey aimed at a systematic study of
the transient sky. The survey camera, mounted on the 48\,in Oschin
Schmidt Telescope at Palomar Observatory (P48), delivers a 7.8 square
degree field of view per image. The survey is now fully operational,
having discovered thousands of transients and hundreds of SNe. See Rau
et al. (2009) for a review of PTF science and observing strategies, and
Law et al. (2009) for performance and technical information concerning
PTF.

One of the key objectives of PTF involves the construction of a large
sample of core-collapse events, for which multicolor optical light
curves and spectra are collected through dedicated follow-up
resources. Such data will allow for a systematic study of the
core-collapse SN population, the first results of which are presented
here.

\subsection{SN Sample}

\begin{figure}
\includegraphics[width=1\textwidth]{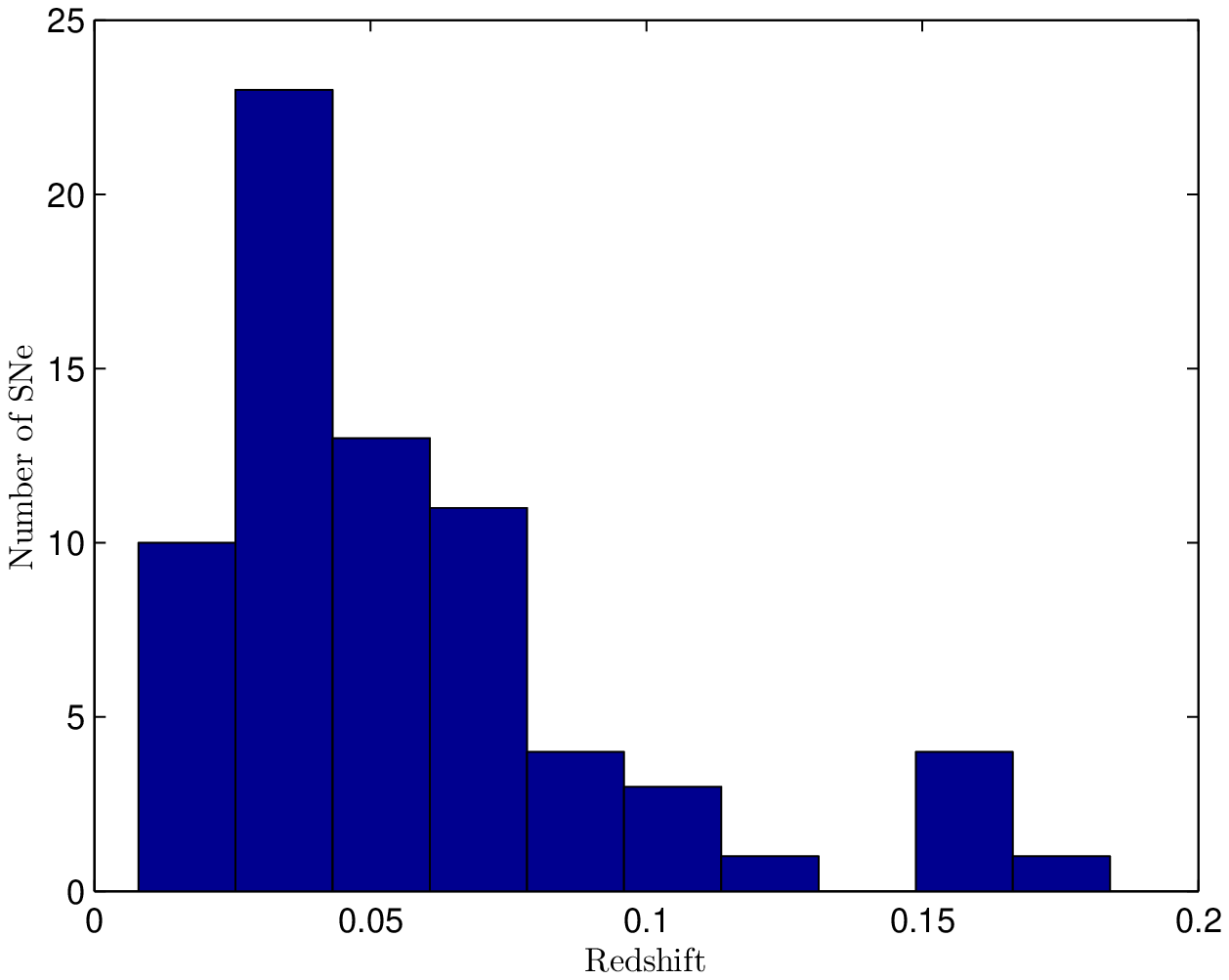}
\caption{The redshift distribution of our sample of 72 SNe.}
\label{figz}
\end{figure}

Our SN sample is comprised of 72 nearby (redshift $z = 0.008$--0.184,
with mean $0.056$ and median $0.046$; Fig. \ref{figz}) core-collapse
SNe discovered by PTF (Table \ref{sntable}; discussed in detail by
Gal-Yam et al. 2010, in preparation) from March 2009 through March
2010. PTF provides an untargeted (i.e., random across the sky, unlike
previous SN surveys which focused on known, generally large galaxies),
homogeneous sample of core-collapse events. The high-cadence PTF
monitoring and its sensitivity largely reduce effects of light-curve
shape and type-specific peak absolute magnitudes in biasing against a
certain variety of SN. While the PTF detection efficiencies have not
yet been fully quantified, any type-specific biases should not depend
on host-galaxy luminosity, and are therefore unlikely to impact our
analysis. The sample analyzed here includes only events
spectroscopically confirmed as core-collapse SNe\footnote{Here we
  exclude new subclasses for which only a few examples have been
  identified, such as pair-production events (Gal-Yam et al. 2009),
  very luminous SNe (Quimby et al. 2009), Type Ibn SNe (Foley et al.
  2007; Pastorello et al. 2008a), SN 2002cx-like events (Li et
  al. 2003), and SN 2002ic-like events (Hamuy et al. 2003).}.

While smaller than the datasets used in previous studies, the nature
of the PTF survey provides a minimally biased homogeneous sample. To
our knowledge it is the largest collection of core-collapse events
ever assembled from a single untargeted survey, rich ($>20\%$) in SNe
from dwarf ($M_r \geq -18$ mag) hosts. Furthermore, detailed subtype
classification has been carried out for all of the SNe in our sample.

\renewcommand{\thefootnote}{\alph{footnote}}

\begin{longtable}{llllrlr}
\caption{PTF Core-Collapse SNe, March 2009 through March 2010\tablenotemark{a}.\label{sntable}}
\\

\hline
\hline
\multicolumn{1}{l}{Name} & \multicolumn{1}{l}{Type} &
\multicolumn{1}{l}{$\alpha$(J2000)} & \multicolumn{1}{l}{$\delta$(J2000)} &
\multicolumn{1}{l}{Host $m_{r}$} & \multicolumn{1}{l}{$z$} &
\multicolumn{1}{l}{Host $M_{r}$} \\ & & & & \multicolumn{1}{l}{[mag]}
& & \multicolumn{1}{l}{[mag]} \\
\hline 
\endfirsthead

\multicolumn{7}{l}%
{{\tablename\ \thetable{}: Continued from previous page}} \\
\hline
\hline
\multicolumn{1}{l}{Name} & \multicolumn{1}{l}{Type} &
\multicolumn{1}{l}{$\alpha$(J2000)} & \multicolumn{1}{l}{$\delta$(J2000)} &
\multicolumn{1}{l}{Host $m_{r}$} & \multicolumn{1}{l}{$z$} &
\multicolumn{1}{l}{Host $M_{r}$} \\ & & & & \multicolumn{1}{l}{[mag]}
& & \multicolumn{1}{l}{[mag]} \\
\hline 
\endhead

\hline
\multicolumn{7}{r}{{\emph{Continued on next page}}} \\
\endfoot

\hline
\endlastfoot

PTF09aux & SN Ic / Ia & 16:09:15.84 & +29:17:36.7 & 16.14 & 0.047 & -20.56\tabularnewline
PTF09awk & SN Ib & 13:37:56.36 & +22:55:04.8 & 17.70 & 0.062 & -19.58\tabularnewline
PTF09axc & SN II & 14:53:13.06 & +22:14:32.2 & 18.12 & 0.115 & -20.61\tabularnewline
PTF09axi & SN II & 14:12:40.82 & +31:04:03.3 & 18.42 & 0.064 & -18.91\tabularnewline
PTF09bce & SN II & 16:35:17.66 & +55:37:59.1 & n/a & 0.023 & n/a\tabularnewline
PTF09bcl & SN II & 18:06:26.78 & +17:51:43.0 & 16.23 & 0.062 & -21.23\tabularnewline
PTF09be & SN II & 14:10:18.53 & +16:53:39.0 & $>$ 21.00 & 0.09 & $>$ -17.13\tabularnewline
PTF09bgf & SN II & 14:41:38.28 & +19:21:43.8 & 17.99 & 0.031 & -17.75\tabularnewline
PTF09bw & SN II & 15:05:02.04 & +48:40:01.9 & 20.28 & 0.15 & -19.04\tabularnewline
PTF09cjq & SN II & 21:16:28.48 & -00:49:39.7 & 13.23 & 0.019 & -21.53\tabularnewline
PTF09ct & SN II & 11:42:13.80 & +10:38:53.9 & 19.38 & 0.15 & -19.99\tabularnewline
PTF09cu & SN II & 13:15:23.14 & +46:25:08.6 & 15.47 & 0.057 & -21.60\tabularnewline
PTF09cvi & SN II & 21:47:09.80 & +08:18:35.6 & 19.40 & 0.03 & -16.41\tabularnewline
PTF09dah & SN IIb  & 22:45:17.05 & +21:49:15.2 & 16.92 & 0.0238 & -18.32\tabularnewline
PTF09dfk & SN Ib & 23:09:13.42 & +07:48:15.4 & 16.90 & 0.016 & -17.45\tabularnewline
PTF09djl & SN II & 16:33:55.94 & +30:14:16.3 & 19.74 & 0.184 & -20.07\tabularnewline
PTF09dra & SN II & 15:48:11.47 & +41:13:28.2 & 16.54 & 0.077 & -21.22\tabularnewline
PTF09due & SN II & 16:26:52.36 & +51:33:23.9 & 13.79 & 0.029 & -21.78\tabularnewline
PTF09dxv & SN IIb  & 23:08:34.73 & +18:56:13.7 & 16.14 & 0.033 & -20.13\tabularnewline
PTF09dzt & SN Ic & 16:03:04.20 & +21:01:47.2 & 17.17 & 0.0874 & -21.09\tabularnewline
PTF09ebq & SN II & 00:14:01.69 & +29:25:58.5 & 15.15 & 0.0235 & -20.01\tabularnewline
PTF09ecm & SN II & 01:06:43.16 & -06:22:40.9 & 16.52 & 0.0285 & -19.35\tabularnewline
PTF09ejz & SN Ic / Ia & 00:55:07.29 & -06:57:05.4 & 16.93 & 0.11 & -21.87\tabularnewline
PTF09fae & SN IIb  & 17:26:20.33 & +72:56:30.6 & 19.78 & 0.067 & -17.72\tabularnewline
PTF09fbf & SN II & 21:20:38.44 & +01:02:52.9 & 14.70 & 0.021 & -20.31\tabularnewline
PTF09fma & SN II & 03:10:23.33 & -09:59:58.0 & 16.17 & 0.031 & -19.74\tabularnewline
PTF09fmk & SN II & 23:57:46.19 & +11:58:45.3 & 16.48 & 0.0631 & -21.02\tabularnewline
PTF09foq & SN II & 05:18:54.83 & +19:10:11.4 & 15.42 & 0.02 & -20.62\tabularnewline
PTF09foy & SN II & 23:17:10.58 & +17:15:03.2 & 16.38 & 0.06 & -20.88\tabularnewline
PTF09fqa & SN II & 22:25:32.33 & +18:59:41.4 & 16.17 & 0.03 & -19.55\tabularnewline
PTF09fsr\tablenotemark{b} & SN Ib & 23:04:52.98 & +12:19:59.0 & 10.74 & 0.007942 & -22.23\tabularnewline
PTF09g & SN II & 15:16:31.48 & +54:27:34.7 & 15.99 & 0.04 & -20.29\tabularnewline
PTF09gof & SN II & 01:22:25.60 & +03:38:08.4 & 17.65 & 0.103 & -20.82\tabularnewline
PTF09gpn & SN II & 03:43:43.26 & -17:08:43.1 & 19.13 & 0.015 & -15.36\tabularnewline
PTF09gtt & SN II & 02:20:37.70 & +02:24:13.2 & 18.68 & 0.041 & -17.72\tabularnewline
PTF09gyp & SN IIb  & 01:58:56.76 & -07:16:56.9 & 21.75 & 0.046 & -14.87\tabularnewline
PTF09hdo & SN II & 00:15:23.20 & +30:43:19.3 & 14.79 & 0.047 & -22.00\tabularnewline
PTF09hzg & SN II & 11:50:57.74 & +21:11:49.4 & 15.43 & 0.028 & -20.12\tabularnewline
PTF09iex & SN II & 12:02:46.86 & +02:24:06.8 & 18.40 & 0.02 & -16.37\tabularnewline
PTF09ige & SN II & 08:55:34.24 & +32:39:57.0 & 17.10 & 0.064 & -20.27\tabularnewline
PTF09igz & SN II & 08:53:56.70 & +33:40:11.5 & 18.77 & 0.086 & -19.27\tabularnewline
PTF09iqd & SN Ic & 02:35:23.23 & +40:17:08.7 & 14.93 & 0.034 & -21.11\tabularnewline
PTF09ism & SN II & 11:44:35.87 & +10:12:43.7 & 17.27 & 0.029 & -18.46\tabularnewline
PTF09ps & SN Ic & 16:14:08.62 & +55:41:41.4 & 19.52 & 0.1065 & -18.96\tabularnewline
PTF09q & SN Ic & 12:24:50.11 & +08:25:58.8 & 16.61 & 0.09 & -21.53\tabularnewline
PTF09r & SN II & 14:18:58.63 & +35:23:16.0 & 17.87 & 0.027 & -17.53\tabularnewline
PTF09sh & SN II & 16:13:58.08 & +39:31:58.1 & 15.35 & 0.0377 & -20.78\tabularnewline
PTF09sk & SN Ic-BL & 13:30:51.15 & +30:20:04.9 & 17.43 & 0.0355 & -18.57\tabularnewline
PTF09t & SN II & 14:15:43.29 & +16:11:59.1 & 16.05 & 0.039 & -20.17\tabularnewline
PTF09tm & SN II & 13:46:55.94 & +61:33:15.6 & 15.33 & 0.035 & -20.66\tabularnewline
PTF09uj & SN II & 14:20:11.15 & +53:33:41.0 & 17.27 & 0.0651 & -20.09\tabularnewline
PTF10bau & SN II & 09:16:21.29 & +17:43:40.2 & 14.03 & 0.026 & -21.33\tabularnewline
PTF10bfz & SN Ic-BL & 12:54:41.27 & +15:24:17.0 & 21.90 & 0.15 & -17.42\tabularnewline
PTF10bgl & SN II & 10:19:04.70 & +46:27:23.3 & 13.63 & 0.03 & -22.00\tabularnewline
PTF10bhu & SN Ic & 12:55:28.44 & +53:34:28.7 & 16.63 & 0.036 & -19.40\tabularnewline
PTF10bip & SN Ic & 12:34:10.52 & +08:21:48.5 & 18.64 & 0.051 & -18.19\tabularnewline
PTF10bzf\tablenotemark{c} & SN Ic-BL & 11:44:02.99 & +55:41:27.6 & 19.05 & 0.0498 & -17.71\tabularnewline
PTF10cd & SN II & 03:00:32.93 & +36:15:25.4 & 19.08 & 0.0455 & -18.00\tabularnewline
PTF10con & SN II & 16:11:39.09 & +00:52:33.3 & 16.78 & 0.033 & -19.36\tabularnewline
PTF10cqh & SN II & 16:10:37.60 & -01:43:00.7 & 15.12 & 0.041 & -21.57\tabularnewline
PTF10cwx & SN II & 12:33:16.53 & -00:03:10.6 & 18.52 & 0.073 & -19.14\tabularnewline
PTF10cxq & SN II & 13:48:19.32 & +13:28:58.8 & 17.55 & 0.047 & -19.11\tabularnewline
PTF10cxx & SN II & 14:47:27.78 & +01:55:03.8 & 15.79 & 0.034 & -20.20\tabularnewline
PTF10czn & SN II & 14:51:16.23 & +15:26:43.6 & 14.97 & 0.045 & -21.63\tabularnewline
PTF10dk & SN II & 05:08:21.54 & +00:12:42.9 & 23.55 & 0.074 & -14.28\tabularnewline
PTF10dvb\tablenotemark{d} & SN II & 17:16:12.25 & 31:47:36.0 & 14.43 & 0.022942 & -20.74\tabularnewline
PTF10hv & SN II & 14:03:56.18 & +54:27:31.1 & 16.93 & 0.0518 & -19.91\tabularnewline
PTF10in & SN IIb  & 07:50:01.24 & +33:06:23.8 & 20.46 & 0.07 & -17.20\tabularnewline
PTF10s & SN II & 10:37:16.30 & +38:06:23.2 & 17.42 & 0.051 & -19.41\tabularnewline
PTF10ts\tablenotemark{e} & SN II & 12:33:56.40 & +13:55:08.3 & 16.84 & 0.046 & -19.80\tabularnewline
PTF10u & SN II & 10:09:58.42 & +46:00:35.2 & 19.83 & 0.15 & -19.45\tabularnewline
PTF10wg & SN II / Ib/c & 03:31:45.78 & -25:24:03.3 & $>$ 21.00 & 0.09 & $>$ -17.12\tabularnewline

\end{longtable}

\footnotetext[1]{Two SNe were omitted: PTF10wg due to a
  classification problem, and PTF09bce because the presence of two
  overlapping galaxies along the line of sight prevented a unique host
  determination. Redshift values have varying accuracy, due to varying resolution and quality of the spectra.}
\footnotetext[2]{SN2009jf}
\footnotetext[3]{SN2010ah}
\footnotetext[4]{SN2010aw}
\footnotetext[5]{SN2009nn}

\renewcommand{\thefootnote}{\arabic{footnote}}

\subsection{Classification}

All SN events were spectroscopically confirmed using either the Low
Resolution Imaging Spectrometer (LRIS; Oke et al. 1995) mounted on the
Keck I 10\,m telescope, the DEep Imaging Multi-Object Spectrograph
(DEIMOS; Faber et al. 2003) mounted on the Keck II 10\,m telescope,
the double spectrograph (DBSP; Oke \& Gunn 1982) mounted on the
Palomar 5\,m telescope, the Intermediate dispersion Spectrograph and
Imaging System (ISIS) mounted on the William Herschel 4.2\,m
telescope, the Kast spectrograph (Miller \& Stone 1993) mounted on the
Shane 3\,m telescope, the Gemini Multi-Object Spectrograph (GMOS; Hook
et al. 2004) mounted on the Gemini North 8\,m telescope, or the
X-Shooter spectrograph (D'Odorico et al. 2006) mounted on the Kueyen
8.2\,m unit of the Very Large Telescope.

\begin{figure}
\includegraphics[width=1\textwidth]{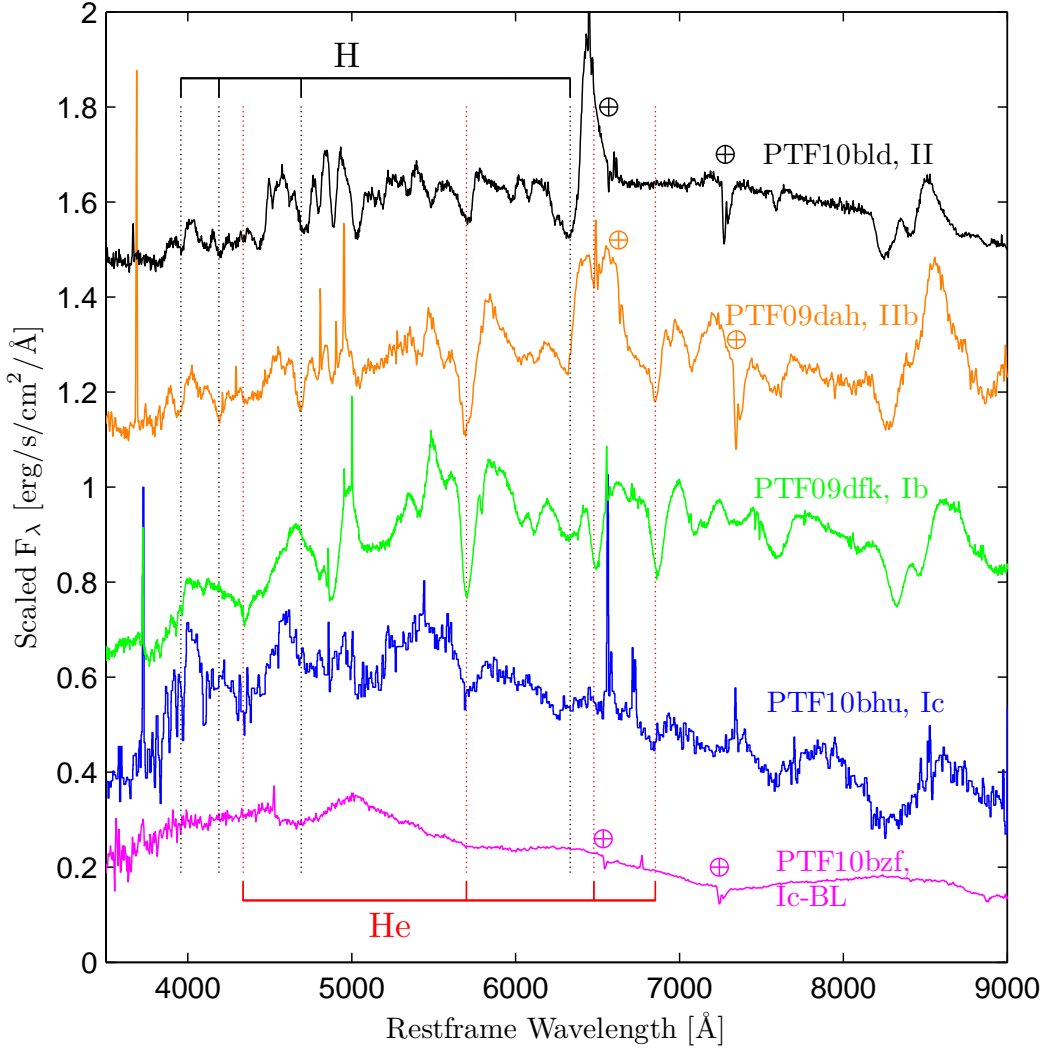}
\caption{Spectroscopic classification. From top to bottom: A normal SN~II (spectrum taken 12 days after peak) exhibits a strong Balmer hydrogen series and lacks prominent
  He~I lines, especially redward of H$\alpha$. In contrast, Type IIb
  events (spectrum taken 30 days after peak) show both hydrogen Balmer lines and the He~I
  absorption series; note the typical absorption ``notch'' from He~I
  $\lambda6678$ superposed on the broad H$\alpha$ emission
  profile. SNe~Ib (spectrum taken approximately 35 days after peak) show strong He~I lines but lack
  obvious H signatures, while in SNe~Ic (spectrum taken about 18 days after peak) the He~I lines
  are weak or absent. The bottom spectrum (spectrum taken some 3 days before peak) shows a
  broad-lined Type Ic event (SN~Ic-BL); note the smooth shape due to
  extreme broadening of the lines (e.g., compare the Ca~II
  near-infrared triplet profile with that in previous spectra). Data
  shown were obtained using Keck+LRIS (PTF10bld and PTF09dfk); P200+DBSP (PTF09dah); 
  WHT+ISIS (PTF10bhu); and Gemini North+GMOS
  (PTF10bzf).  The main telluric absorption features are marked.}
\label{figspec}
\end{figure}

Events deemed most likely to be core-collapse SNe were classified
(Gal-Yam et al.  2010, in preparation; Fig. \ref{figspec}) as Type II
(prominent H lines), IIb (H lines as well as prominent He lines), Ib
(no H; strong He lines), and Ic (no H or He lines). Both visual and
automatic classification, using Superfit (Howell et al. 2005) and SNID
(Blondin et al. 2007), were carried out. If a better match was found
to SN~Ic-BL templates (mainly SN 1998bw and SN 2002ap), the event was
classified as a SN~Ic-BL. Note that for the purpose of this paper, we
do not distinguish between the different SN~II subtypes (IIn,
displaying narrow and intermediate-width components; IIP/L, based on
light-curve shape) except for Type IIb. This is in accord with direct
detection of H-rich progenitors for SNe~IIP and IIn (Gal-Yam et
al. 2007; Gal-Yam \& Leonard 2009; Smartt 2009), while analysis of the
recent SN~IIb 2008ax (Crockett et al. 2008a; Pastorello et al. 2008b;
Chornock et al. 2010) may indicate a stripped progenitor (Crockett et
al. 2008a). We consider Type II (H-rich), IIb (H-poor), Ib (H-less),
and Ic (H- and He-less) as resulting from a progression of progenitors
with increasingly intensive mass loss.

The spectra of two events (PTF09aux and PTF09ejz) were consistent with
both SN~Ia and SN~Ic classifications. For statistical purposes, each
was counted as half a Ic event. For one event (PTF10wg) the type of
core-collapse SN could not be reliably determined, and so it was
omitted from the study.

\subsection{Galaxy Luminosities}

Galaxy luminosities were taken from the Sloan Digital Sky Survey
(SDSS) Data Release 7 (Abazajian et al. 2009), when available. For
those hosts without SDSS coverage, pre-explosion images taken by the
PTF camera on the P48 were constructed, and $r$-band magnitudes were
measured using aperture photometry, calibrated against USNO-B1
red-filter images (Monet et al. 2003). One SN (PTF09fsr) host
magnitude was retrieved from the NASA/IPAC Extragalactic Database
(NED)\footnote{http://nedwww.ipac.caltech.edu/.}. Spectroscopic
redshifts were used to calculate absolute magnitudes assuming a
cosmological model with $H_0 =
70\,\textrm{km\,s}^{-1}\,\textrm{Mpc}^{-1}$, $\Omega_{m}=0.3$, and
$\Omega_{\Lambda}=0.7$. Galactic extinction was removed using the
Schlegel et al. (1998) maps via NED. K-corrections are negligible for
the typical low redshifts of the sample and were ignored. We adopt a
criterion of $M_r = -18$ mag to distinguish between giant galaxies
($M_r < -18$ mag) and dwarf galaxies ($M_r \geq -18$ mag), based on
the observed luminosity distribution (Fig. \ref{fighostmag}). This
value coincides with physically motivated metallicity thresholds for
long-duration GRB occurrence presented by Woosley \& Heger (2006).

The host of PTF09be was not detected in pre-explosion images, but the
upper limit on its luminosity indicates that it is a very faint dwarf
galaxy. Several host galaxies\footnote{Those of PTF09t, PTF09cjq,
  PTF09due, PTF09fbf, PTF09hzg, PTF10bau, PTF10bgl, PTF10czn and PTF10dvb.}
subtend a large angular size and were fragmented by the SDSS
photometry pipeline. We derived a lower luminosity limit using the
cataloged brightness of their core, which places all 9 of them in the
giant-galaxy class. Finally, one event (PTF09bce) occurred in an area
with two possible host galaxies near the line of sight, overlapping
each other; it was therefore omitted from the study.

\subsection{Galaxy Metallicities}

In order to enable a comparison with other studies, we must translate
our $r$-band luminosities to metallicity values, preferably in solar
metallicity units. The procedure for doing this is not
straightforward, as it depends on the adopted metallicity scale and
the adopted luminosity vs. metallicity relation. We obtained
metallicity estimates in the following manner. Using the SDSS $g$- and
$r$-band magnitudes of 39 host galaxies for which these data are
available, together with the transformation given by Smith et
al. (2002), we find the host $B$-band magnitudes. We then use the
luminosity vs. metallicity relation presented by Tremonti et
al. (2004) to find the host metallicities in terms of
$\log\left(\textrm{O/H}\right)$. Adopting a solar metallicity of
$12+\log\left(\textrm{O/H}\right)=8.9$ (Delahaye \& Pinsonneault
2006), we can then translate our deduced metallicities to the solar
scale. At this point, we have metallicity values for 39 host galaxies
to which we fit the
relation \[\log\left(Z/Z_{\odot}\right)=-0.161\left(\pm0.012\right)
M_{r}-3.36\left(\pm0.23\right)\]
(with root-mean square residuals of 0.048; the uncertainties of the
fit parameters designate a 95\% confidence interval), and use it to
estimate the metallicity of all the host galaxies for which $r$-band
magnitudes are available.

\section{Results}

\begin{figure}
\includegraphics[width=1\textwidth]{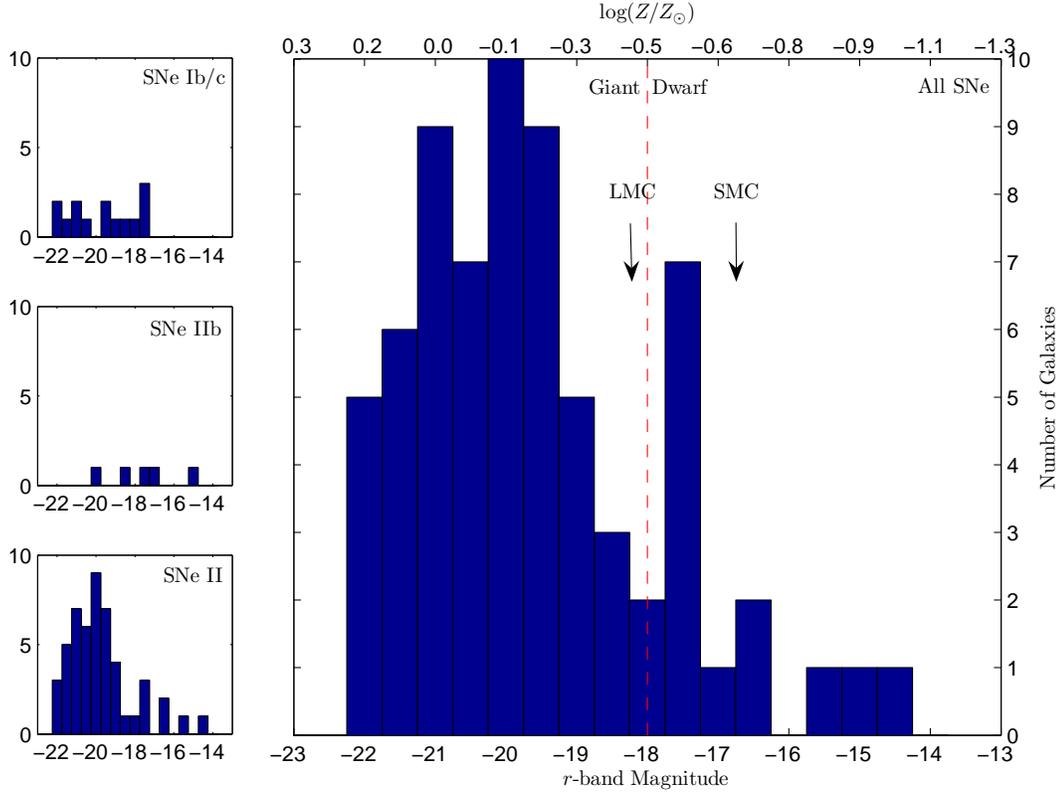}
\caption{The $r$-band absolute magnitudes of the host galaxies of our
  72 SNe, except two for which only upper limits are available,
  and one for which a host could not be determined. The dwarf/giant
  cut we adopt is represented by the dashed line, and metallicity
  estimates are noted in the upper axis. The $r$-band magnitudes of
  the Large and Small Magellanic Clouds are also noted, converted from
  NED using the Smith et al. (2002) equations with distance moduli
  adopted from Keller \& Wood (2006).  Host-magnitude histograms for
  different SN types are shown at left. SNe~Ib/c are seen to be
  concentrated in luminous hosts. SNe~IIb are located in the
  intermediate hosts and the SN~II population extends toward the
  dimmest galaxies, as expected from metallicity considerations. An
  excess of SNe~II in giant hosts compared to dwarf hosts is also
  evident.}
\label{fighostmag}
\end{figure}

\begin{figure}
\includegraphics[width=1\textwidth]{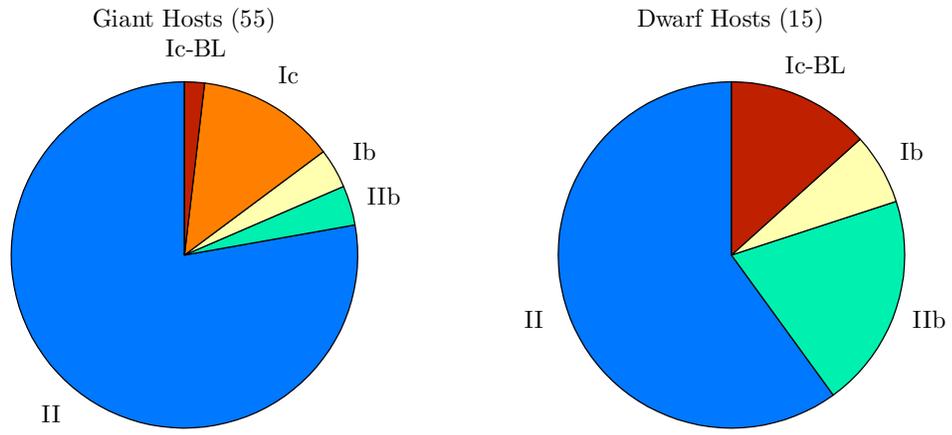}
\caption{Division of SN types in 55 giant galaxies (left) and 15 dwarf
  galaxies (right). Note the overabundance of SNe~IIb and SNe~Ic-BL in
  dwarf hosts and the lack of ``normal'' SNe~Ic there.}
\label{figpie}
\end{figure}

A histogram of host $r$-band absolute magnitudes is presented in
Figure \ref{fighostmag}. A bright population centered around $M_r =
-20$ mag can be identified, as well as a faint tail extending to $M_r
= -14$ mag. Our choice of $M_r = -18$ mag as a limiting value between
giant and dwarf galaxies is marked. The number of SNe of each type
occurring in giant and dwarf galaxies is listed in Table
\ref{typestable} and shown in Figure \ref{figpie}. Shifting the
dwarf/giant cut to $M_r = -18.5$ mag increases the dwarf-galaxy sample
by three: one SN~II, one SN~IIb (leaving only one SN~IIb in giant
hosts), and one SN~Ic (making it the only one in a dwarf host under
this definition).

\renewcommand{\thefootnote}{\alph{footnote}}

\begin{table}
\caption{Division of SN types for each host-galaxy class\tablenotemark{f}.\label{typestable}}
\begin{tabular}{llll}
\hline 
\hline 
SN Type & No. in Giants (Fraction) & No. in Dwarfs (Fraction) & Likelihood \tabularnewline
\hline
II & 42 ($0.78_{-0.07}^{+0.06}$) & 9 ($0.60_{-0.16}^{+0.14}$) & $6\%$ \tabularnewline
IIb & 2 ($0.04_{-0.02}^{+0.05}$) & 3 ($0.20_{-0.11}^{+0.15}$) & $2\%$ \tabularnewline
Ib & 2 ($0.04_{-0.02}^{+0.05}$) & 1 ($0.07_{-0.06}^{+0.13}$) & $29\%$ \tabularnewline
Ic & 7 ($0.13_{-0.05}^{+0.06}$) & 0 ($0.00_{-0.00}^{+0.05}$) & $7\%$ \tabularnewline
Ic-BL & 1 ($0.02_{-0.02}^{+0.04}$) & 2 ($0.13_{-0.08}^{+0.15}$) & $4\%$ \tabularnewline
\hline 
Galaxies & 55 & 15 & \tabularnewline
\hline
\end{tabular}
\end{table}

\footnotetext[6]{Two
  giant-host SNe~Ic were counted as half each, due to the possibility
  of each being a SN~Ia. Two out of our sample of 72 SNe were omitted
  (one due to a classification problem and one because of ambiguous
  host determination). Fractions are shown with $1\sigma$
  uncertainties, calculated using the Clopper-Pearson method (see
  Gehrels 1986). The likelihood of both fractions of each SN type
  arising from the same distribution is displayed, calculated by the
  method described in $\S~4$ for the likelihood of the SN~IIb and
  SN~Ic-BL overabundances arising by chance.}

\renewcommand{\thefootnote}{\arabic{footnote}}

\subsection{Caveats}

There are a number of caveats to keep in mind when considering our
results.

(1) Our analysis is currently limited by the small number of
events. This precludes some comparisons with other studies and
theoretical predictions, but does allow for significant trends to be
identified, as stated below.

(2) The difficulty of identifying SNe in the central core of giant
galaxies may induce a bias against SNe~Ib/c, which are more
concentrated in those regions (Anderson \& James 2009). However,
Botticella et al. (2008) find that $\sim 50$\% of galactic core
transients are not SNe. Furthermore, in PTF the difficulty arises only
in the central $\sim1''$ of the galaxies (which have much larger
typical sizes), so we expect this effect to be negligible (indeed,
most nuclear PTF events are active galactic nuclei).

(3) In some cases, SNe can be confused with other transients. For this
reason, the spectra of all SN candidates were checked by eye as well
as using Superfit (Howell et al. 2005) and SNID (Blondin et
al. 2007). We retained those candidates which are most likely SNe.

(4) Some SNe may have undergone evolution from one subclass to another
(for example, from Type II to Type IIb), which could have escaped our
notice due to limited spectroscopic coverage.  A detailed discussion
of this issue will be given in a companion paper (Gal-Yam et al. 2010,
in preparation); however this uncertainty exists for SNe regardless of
host luminosity, and is thus not expected to bias our results
systematically.

(5) Our derived metallicity values are influenced by the chosen method
and adopted parameters. In addition, some nearby GRB hosts have been
found to lie below the luminosity vs. metallicity relation (i.e., have
lower metallicities than their luminosities suggest; Kewley et
al. 2007); moreover, while many SNe~Ic-BL hosts do lie along the
luminosity vs. metallicity relation, some outliers have been observed
(Modjaz et al. 2008). Using direct host-galaxy measurements could thus
be more insightful. However, because we do not state any
metallicity-specific conclusions in this work, we leave a more
detailed treatment of this issue to a future paper.

(6) Metallicity varies within galaxies, while here we deal with global
metallicity and luminosity values. Indeed, a location study of our SN
sample would complement these results.

\section{Discussion}

While small, the number of core-collapse SNe discovered in dwarf
galaxies is higher than expected. Using rough scaling relations to
compare PTF with the Pan-STARRS1 predictions presented by Young et
al. (2008), it would appear that the number of such SNe is larger than
predicted by a factor of a few. Two of these events, PTF10bfz and
PTF10bzf (SN 2010ah; Ofek et al. 2010), are SNe~Ic-BL having no
obvious association with a GRB. Along with the recent discovery of SN
2007bg (Moretti et al. 2007; Young et al. 2009) as a dwarf-host
(Prieto et al. 2008) SN~Ic-BL without an associated
GRB\footnote{Neither SN 2007bg nor PTF10bzf show radio emissions
  indicative of the relativistic outflows present in GRBs (Soderberg
  2009; Chomiuk \& Soderberg 2010).}, these events suggest that not
all SNe~Ic-BL in low-metallicity galaxies are associated with
long-duration GRBs (cf. Young et al. 2008; Modjaz et
al. 2008). However, without direct metallicity measurements, we cannot
determine whether they violate the dividing line presented by Modjaz
et al. (2008), as these hosts may be outliers to the metallicity
vs. luminosity relation.

We find that the SN~Ib/c to SN~II number ratio is
$N(\textrm{Ib/c})/N(\textrm{II})=0.23_{-0.08}^{+0.11}$ in giant
galaxies and $N(\textrm{Ib/c})/N(\textrm{II})=0.25_{-0.15}^{+0.3}$ in
dwarfs ($1\sigma$ uncertainties\footnote{These confidence levels have
  been calculated using the Clopper-Pearson method, discussed, for
  example, by Gehrels (1986).}). While this does not recover the
reduction in the SN~Ib/c fraction in dwarf or low-luminosity hosts
reported by Prantzos \& Boissier (2003) and Prieto et al. (2008), the
difference is not significant. Theoretical studies (Fryer et al. 2007)
predict a fraction of $N(\textrm{Ib/c})/N(\textrm{II}) \lesssim 0.1$
in dwarf hosts. Distinguishing between SN~Ib/c to SN~II ratios of 0.25
and 0.1 at the 95\% confidence level requires a dwarf-host SN sample
size of $\geq57$ (assuming binomial statistics), which is expected to
be attained by PTF within the next few years. The data to be gathered
by PTF (as well as other surveys) will thus allow strong conclusions
to be drawn in the near future.

Our distinction among SNe~Ib, Ic, and Ic-BL subtypes, as well as
separating Types II and IIb, demonstrates the power of a detailed
spectroscopic classification. We find that the only SNe~Ib/c in dwarf
galaxies are either Ib or Ic-BL. ``Normal'' SNe~Ic, abundant in large
galaxies, are not present, or are rare, in dwarf hosts. This
is in agreement with past trends: all SNe~Ib/c in dwarfs studied by
Prieto et al. (2008) were either SNe~Ib (SN 2005hm) or SNe~Ic-BL (SN
2006qk, SN 2007I, and SN 2007bg), while all GRB-associated SNe in
dwarf hosts are also Ic-BL events (Woosley \& Bloom
2006). Discoveries from the ROTSE-III surveys (Quimby 2006; Yuan et
al. 2007) may include one normal SN~Ic hosted by a galaxy meeting our
definition of a dwarf (SN 2004gk; Quimby et al. 2004), but the sample
otherwise shows trends similar to the PTF results.  

We further identify an overabundance of SNe~IIb and SNe~Ic-BL in dwarf
galaxies. Assuming the fraction $p$ of SNe~IIb is independent of host
luminosity, we can calculate the probability of finding $\leq 2$ SNe~IIb
in 55 bright ($M_r < -18$ mag) hosts, and $\geq 3$ SNe~IIb in 15 dwarf
($M_r \geq -18$ mag) hosts. This probability as a function of $p$ is
given by
\[\sum_{x=0}^{2}\left(\begin{array}{c}55\\x\end{array}\right)p^{x}
\left(1-p\right)^{55-x} \sum_{y=3}^{15}\left(\begin{array}{c}15\\y
\end{array}\right)p^{y}\left(1-p\right)^{15-y}.\]
\noindent
Varying $p$ between 0 and 1, we find that this probability is
$<5\%$. A similar result is found for the overabundance of SNe~Ic-BL
in dwarf hosts. It is therefore unlikely that the observed trends are
produced by chance from the same underlying population.

Our data suggest that ``normal'' SNe~Ic occurring in giant galaxies
are replaced by IIb and Ic-BL events in dwarf galaxies. This seems to
confirm both trends suggested by previous location and
population-ratio studies, implying two distinct mechanisms at
work. Metallicity might be the main process responsible for the
evolution of stars which explode as ``normal'' SNe~Ic. In lower
metallicity environments, stars that would otherwise explode as SNe~Ic
end their lives as Ib or IIb events, while many SN~Ib progenitors are
also driven toward Type IIb explosions following a decrease in their
mass loss. However, an additional process which allows for strong
stripping in low-metallicity environments must be present to explain
the SNe~Ic-BL and long GRB events observed in dwarf hosts. Binarity
and rapid stellar rotation are possible candidates for such a
stripping mechanism. Finally, we note an apparent excess of stripped
(IIb+Ib+Ic+Ic-BL) SNe in dwarf galaxies, compared to ``normal'' SNe~II
(see Figures \ref{fighostmag} and \ref{figpie}). If this is indeed the case, it is
unlikely to be caused by mass-loss mechanisms, which are expected to
be stronger in high-metallicity environments. A top-heavy initial mass
function (IMF) in dwarf galaxies might be required to explain this
effect (see also Hakobyan et al. 2009; Habergham et al. 2010). Such
an IMF is also consistent with pair-production SNe (Gal-Yam et al. 2009)
and highly luminous SNe (Quimby et al. 2009), which are associated
with very massive progenitors, occurring only in dwarf hosts.

Our analysis is limited by small-number statistics. Continued
observations by PTF are expected to rapidly increase the sample size
during the coming few years. An attractive complementary study is to
analyze the locations of our SNe within their host galaxies (e.g., as
done by Fruchter et al. 2006 and Kelly et al. 2008 on similar-sized
samples). A combined analysis should provide a powerful constraint on
the evolution and mass-loss properties of massive stars as a function
of their metallicity. The distance and angular size of our host sample
require the high spatial resolution of the {\it Hubble Space
  Telescope} to carry out this test.

\section{Summary}

We have analyzed the initial compilation of core-collapse SNe
discovered by the PTF during its first post-commissioning year, which
comprises the largest homogeneous collection of core-collapse events
from a single untargeted survey ever assembled. Our data indicate an
absence of ``normal'' SNe~Ic in dwarf galaxies offset by an excess of
SNe~IIb and SNe~Ic-BL. We deduce that a metallicity-dependent process
might be responsible for turning ``failed SNe~Ic'' into SNe~Ib and
IIb, while a second process forms the progenitors of SNe~Ic-BL and, by
association, perhaps also of long-duration GRBs in dwarf galaxies.

Our sample of SNe should increase substantially in the coming
years. Ultimately, we plan to analyze the distribution of the
different types of SNe as a function of host-galaxy luminosity,
without having to impose an artificial dwarf/giant cut. In addition,
high-resolution imaging of host galaxies will enable location studies
to be conducted on this SN sample, including (for the first time) a
large fraction of dwarf hosts. Such analyses will continue to expand
the use of SNe in dwarf galaxies as a powerful tool in the study of
massive stars.

\bigskip
\medskip

The Weizmann Institute PTF partnership is supported by the Israeli
Science Foundation via grants to A.G. Collaborative work between
A.G. and S.R.K. is supported by the US-Israel Binational Science
Foundation. A.G. further acknowledges support from the EU FP7 Marie
Curie program via an IRG fellowship, the Benoziyo Center for
Astrophysics, a Minerva grant, and the Peter and Patricia Gruber
Awards.  P.E.N. is supported by the US Department of Energy Scientific
Discovery through Advanced Computing program under contract
DE-FG02-06ER06-04.  M.S. acknowledges support from the Royal Society;
M.S. and A.G. are also grateful for a Weizmann-UK Making Connections
grant.  A.V.F. and S.B.C. acknowledge generous support from Gary and
Cynthia Bengier, the Richard and Rhoda Goldman Fund, US National
Science Foundation grant AST-0908886, and the TABASGO Foundation.
J.S.B. was partially supported by a SciDAC grant from the
U.S. Department of Energy and a grant from the National Science
Foundation (award 0941742).  The National Energy Research Scientific
Computing Center, which is supported by the Office of Science of the
U.S. Department of Energy under Contract No. DE-AC02-05CH11231,
provided staff, computational resources, and data storage for this
project.

The WHT is operated on the island of La Palma by the Isaac Newton
Group in the Spanish Observatorio del Roque de los Muchachos of the
Instituto de Astrof\'{\i}sica de Canarias.  Some of the data presented
herein were obtained at the W. M. Keck Observatory, which is operated
as a scientific partnership among the California Institute of
Technology, the University of California, and NASA; the observatory
was made possible by the generous financial support of the W. M. Keck
Foundation. We are grateful to the staff of the Keck, Lick, Palomar,
Roque de los Muchachos, VLT, and Gemini Observatories for their
assistance.  We thank Chris Lidman for processing the X-Shooter data
used for the classification of PTF10bau.

\end{document}